%% file: Shirokov_A.tex
\documentclass[
aps,%
11pt,%
final,%
titlepage,%
oneside,%
twocolumn,%
nobibnotes,%
nofootinbib,%
superscriptaddress,%
showkeys,%
centertags]%
{revtex4}

\usepackage{amsmath}
\usepackage{graphicx}
\usepackage[colorlinks=true,linkcolor=blue,citecolor=blue,urlcolor=blue]{hyperref}

\setcitestyle{authoryear}

\input{cmd_arXiv.tex}

\usepackage{etoolbox}

 \makeatletter
 \renewcommand\@biblabel[1]{\hspace*{\labelwidth}}
 \apptocmd{\NAT@thebibliography}{\setlength\itemindent{15pt}}{}{}
 \makeatother

\usepackage{setspace}

\begin{document}

\singlespacing

\title{ THESEUS--BTA cosmological crucial tests using \\
Multimessenger Gamma-Ray Bursts observations}

\author{\firstname{S.~I.}~\surname{Shirokov}}
\email{arhath.sis@yandex.ru}
\affiliation{SPb Branch of Special Astrophysical Observatory of Russian Academy of Sciences, 65 Pulkovskoye shosse, St. Petersburg, 196140, Russia}
\author{\firstname{Ilia~V.}~\surname{Sokolov}}
\email{ilia.v.sokolov@gmail.com}
\affiliation{Institute of Astronomy of the Russian Academy of Sciences, Pyatnitskaya st., 4, Moscow, 119017, Russia}
\author{\firstname{V.~V.}~\surname{Vlasyuk}}
\affiliation{Special Astrophysical Observatory,
Nizhnij Arkhyz, Zelenchukskiy region, Karachai-Cherkessian Republic, 369167, Russia}
\author{\firstname{L.}~\surname{Amati}}
\affiliation{INAF, Istituto di Astrofisica Spaziale e Fisica Cosmica, Bologna, Via Gobetti 101, I-40129 Bologna, Italy}
\author{\firstname{V.~V.}~\surname{Sokolov}}
\affiliation{Special Astrophysical Observatory,
Nizhnij Arkhyz, Zelenchukskiy region, Karachai-Cherkessian Republic, 369167, Russia}
\author{\firstname{Yu.~V.}~\surname{Baryshev}}
\affiliation{Saint Petersburg State University,
Universitetskij pr. 28, Staryj Peterhoff, St. Petersburg, 198504, Russia}
\begin{abstract} 
Modern Multimessenger Astronomy is a powerful instrument for performing cosmological crucial tests of the Standard  Cosmological Model in the wide redshift interval up to $z \sim 10$. This is principally important for discussion 
related to discrepancies between local and global measurements of cosmological parameters.
We present a review of multimessenger gamma-ray burst observations currently conducted and planed for THESEUS--BTA cooperative program.
Such observations give a unique opportunity to test the fundamental foundations of cosmological models: gravitation theory; 
cosmological principle of homogeneity and isotropy of large-scale distribution of matter; and  space expansion paradigm. Important role of various selection effects leading to systematic distortions of true cosmological relations is discussed.
\end{abstract}

\maketitle

\section{Introduction}

Recent discussion of the standard Lambda Cold non-baryonic Dark Matter ($\Lambda$CDM) model uncovered \textit{``possible crisis for cosmology'' } 
\citep{Baryshev2015,Handley2019,Lin2019,Verde2019,DiValentino2019,DiValentino2020,Riess2020}, which demonstrated that the large-scale cosmological physics contains several principle uncertainties. Among them: the absence of crucial decision on  closed-flat-open (curvature parameter $k=+1,\,0,$ and $-1$) geometry of the universe \citep{DiValentino2019, DiValentino2020, Handley2019}, the nature and value of the totally dominating dark energy and non-baryonic   dark matter, $\Omega_{de},\,w,$ and $\Omega_m$,
\citep{Makarov2011,Colin2019,Demianski2019,DiValentino2020}
the difference between local and global values for the Hubble constant $H_0$ 
\citep{Tully2016,Riess2018b,Lin2019,Verde2019,Riess2020}
are especially worrying problems. The currently observed discordances may indicate the need for new physics and possibly point to drastic changes in the $\Lambda$CDM scenario 
\citep{Lin2019,Verde2019,DiValentino2019,DiValentino2020,Riess2020}.

Modern theoretical and experimental physics  also tests foundations of the Standard Cosmological Model (SCM). In particular, the modified theories of gravity change the study of cosmic structure formation~\citep{Clifton2012, Ishak2018, Bartelmann2019, Slosar2019} (review~\citet{Ishak2018} contains 900 references). 

This new situation in cosmology stimulates deep testing of the fundamental physical laws at micro and macro scales simultaneously.
Modern physics uses the observable Universe as a part of physical laboratory, where the fundamental physical laws  must be tested. 
Such basic theoretical assumptions as: constancy of the fundamental constants $c,\,G,\,m_p,$ and $m_e$, the Lorentz invariance, the equivalence principle, the quantum principles of the gravity theory, are now being investigated  by modern theoretical physics~\citep{Uzan2003,Rubakov2008,deRham2014} and by contemporary astrophysical observations~\citep{Uzan2010,Clifton2012,deRham2017,Giddings2017,Ishak2018,Cardoso2019}.

In fact, in the beginning of the 21st century a New Cosmology emerges and a new set of questions arises. In particular, the famous Turner's list of new cosmological problems contains the following puzzles:  What is the physics of underlying inflation?  How was the baryon asymmetry produced? What is the nature of the non-baryonic dark matter particles? Why is the composition of our Universe so ``absurd'' relative to the lab physics? What is the nature of the dark energy? Answering these questions will reveal deep connections between fundamental physics and cosmology: \emph{``There may even be some big surprises -- time variation of the constants or a new theory of gravity that eliminates the need for dark matter and dark energy''}~\citep{Turner2002}.


The visible matter of the Universe, the part, which we  can actually observe, is a surprisingly small (about 0.5\%) piece of the predicted matter content and this looks like an ``Absurd Universe''~\citep{Turner2003}. What is more, about 95\% of the cosmological matter density, which determine the dynamics of the whole Universe has an unknown physical nature. Turner emphasized that: \emph{``modern SCM predicts with high precision the values for dark energy and non-baryonic cold dark matter, but we have to make sense to all this''}~\citep{Turner2002}.

Current multimessenger astronomy, including observations of gamma-ray bursts (GRBs), offers new capabilities for cosmological tests,  especially in view of the forthcoming mission Transient High Energy Sky and Early Universe Surveyor (THESEUS) \citep{Amati2018,Strata2018}. Preceding reviews of the GRB cosmology were given by \citet{Petrosian2009,Wang2015}. 
We show that an important contribution to GRB cosmology belongs to cooperative  optical observations using the 6m SAO BTA telescope facilities (\citep{Vlasyuk2018}). 

In our paper we adopt the Sandage's ``practical cosmology'' approach 
\citep{Sandage1997,Sandage1995a}, 
started by \citet{Hubble1937,Hubble1935}.
According to it one should  test the initial principles of cosmological models by using new astronomical facilities.
We review application of GRB multimessenger observations 
to classical cosmological tests described by \citet{Baryshev2012}.
Tests which can probe the underlying basic principles of SCM are in the focus. 
In particular, we consider  GRB multimessenger data for testing a gravity theory,  Cosmological Principle and space expansion paradigm.

In Sec.2 we list the SCM basic principles to be tested by THESEUS--BTA facilities. Such tests can strengthen the validity of the SCM foundation or point to limitation of its application. 
Gravity theory in strong field regime and its testing with GRB observations is considered in Sec.3.
Sec.4 describes GRBs in application to the Cosmological Principle testing.
In Sec.5 we discuss 
GRB studies as an instrument for testing the space expansion paradigm, including the Hubble Diagram and 
Wilson's time delay test.
Conclusions are given in Sec.6.

\section{Testing the SCM  basis by multimessenger GRB observations}

The success of the SCM in explaining the main cosmologically important observations is generally recognized~\citep{Peebles1993,Peacock1999,Baryshev2012}. 
Fundamental physical basis of the SCM contains the following theoretical assumptions:
\begin{itemize}
\item[$\bullet$] 
General Relativity Theory (GRT) -- all gravity phenomena can be described by the metric tensor $g_{ik}$ of the Riemannian space $\mathcal{R}$.  
\item[$\bullet$]
 Einstein's Cosmological Principle --
the strict mathematical homogeneity for the dynamically important matter, i.e. 
$\rho (\Vec{r},t) = \rho (t) $, 
$\,p (\Vec{r},t) = p (t) $,
$\,\,\,g_{ik} = g_{ik}(t)$ (on all scales $r$ the matter density $\rho$, the total pressure $p$ and the space metric $g_{ik}$ are
functions of time only). 
\item[$\bullet$]
Expanding space paradigm --  time dependent distances between galaxies
$r(t)=S(t)\chi$, where $S(t)$ is the scale factor and $\chi$ is the comoving distance -- according to which the observed cosmological redshift is interpreted as the Lema\^{i}tre effect of the space stretching (not the Doppler effect).
\end{itemize}

At the beginning of  the 21st century  professional cosmological community began to discuss 
the validity and possibilities for testing  these three conceptual ``pillars'' of the SCM \citet{Baryshev2015}. 

The abilities of 
gamma-ray, X-ray, IR instrumentation onboard THESEUS 
\citep{Amati2018,Strata2018}
accompanying with GRB studies at the 6m BTA SAO observations
\citep{Vlasyuk2018,Sokolov2018a}
and other optical ground-based telescopes can play a crucial role in testing basic principles of the SCM.

Modern achievements of theoretical physics,  especially  different modifications of general relativity and  quantum aspects of gravitation theory, require wider observational testing  of the basic SCM principles.
Nowadays physicists consider the observable Universe as a part of ``cosmic laboratory'', where main physical fundamental laws must be tested in wider redshift interval and with increasing accuracy. The THESEUS GRB observations will provide such an opportunity for redshifts up to $z \sim 10 $.
In particular,  the Cosmological Principle, the general relativity and its modifications, the space expansion paradigm, must be tested by new observations.

We consider 
the following a strategy:
\begin{itemize}
    \item gamma-to-IR observations of the massive core collapse supernovae (long GRBs) and the merging of binary neutron stars (Short GRBs) to test the gravity theory;
    \item testing the Cosmological Principle of homogeneity and isotropy in studies of the spatial distribution of GRBs host galaxies and sight-line distribution of galaxies towards GRBs;
    \item constructing the high-redshift GRB Hubble diagram and comparison of time dilation in GRB pulses, GRB afterglow and core-collapse SN light curves to test the expanding space paradigm.
\end{itemize}
These tests are of the fundamental importance for developing an adequate  cosmological model that includes modern multimessenger observations of  GRBs.

\section{Testing gravity theory by GRB observations}
The most important basic assumption of $\Lambda$CDM is the general relativity theory. It has been successfully tested in the weak gravity conditions. 
But nowadays theoretical physics suggests various new possibilities for the modification of GRT (\citep[see][]{Giddings2017,Ishak2018,Cardoso2019}.
For this reason, crucial cosmological tests including the modern alternative gravitation theories in strong-gravity regime are needed.
The cosmological model is a solution of the gravitational field equations for the case of a cosmologically large-scale distribution of matter.

\subsection{The quest for unification of gravity with other fundamental forces}

Being a prototype of the geometrical approach to gravitation GRT is a non-quantum theory, so it does not obey the quantum principles of modern physics. The most challenging problem of modern theoretical physics is to construct the quantum theory  of gravitation which is united with other fundamental quantum interactions -- strong, weak, electromagnetic \citep{AmelinoCamelia2000,Hawking2014,Wilczek2015,Giddings2017}.

In general, there are two  alternative approaches for inclusion of gravitation to unified theory: 1) modification of existing theories of fundamental interactions to include them into curved geometry,
or 2) development of a quantum field gravity theory of gravity based on general principles with other fundamental physical interaction (Minkowski space-time, positive localizable field energy density, energy-momentum conservation, uncertainty principle, quanta of gravity field energy).

The first approach is based on modification of geometrical description by quantization of the curved space-time \citep{Rovelli2004}. 
However, there is an important obstacle for unification of fundamental forces with the geometrical gravitation theory: the conceptual basis of GRT is principally different from the Elementary Particle Standard Model (EPSM).
Gravity in the framework of GRT is not a force and there is no generally covariant Energy-Momentum Tensor of the gravity field (the problem of psudotensor and localization of the field energy formulated by
\citet{Landau1971}). 

The second approach is based on developing a (non-metric) field gravity theory using the modern quantum field theory (QFT).
The success of the Standard Model of electromagnetic, weak and strong interactions was achieved on the way of unification of the fundamental physical
forces in the frame of the QFT. Now it has reached a
respectable status as an accurate and well-studied description of sub-atomic forces and particles, though some conceptual and technical problems remain
to be solved \citep{Wilczek2015}. 

It is expected that the future ``Core Theory'' of physics will unify all fundamental forces (electromagnetic, weak, strong and gravitational) and also deliver unification of forces (bosons) and substances (fermions) via transformations of supersymmetry \citep{Wilczek2015}.

\citet{Feynman1995} considered the construction of quantum field theory of gravity as the symmetric second rank tensor field in Minkowski space, based on common principles with other fundamental forces. A development of Feynman's approach was done in 
\citet{Sokolov1980,Baryshev2012,Sokolov2015,Sokolov2016,Baryshev2017,Sokolov2019,Baryshev2019}, where new predictions were considered for structure of the relativistic compact objects (RCO) and for cosmological solution of the gravity field equations.

Both modified geometrical and quantum field approaches should be studied more carefully and tested by astrophysical observations of RCOs  and cosmological processes. 

\subsection{GRB observations for testing strong gravity effects}
The space and ground-based multimessenger observations of GRBs can be used to get important restrictions on possible gravitation theories. 

We initiate an international observational program on monitoring of GRBs detected by Swift, Fermi, INTEGRAL, Lomonosov and other space missions. 
The program  aims at searching for optical/electromagnetic counterparts connected with GRBs, neutrino sources and gravitational wave (GW) events detected by the Laser Interferometer Gravitational-wave Observatory (LIGO) and Virgo experiment. These observations present an international collaboration developing the future THESEUS  space mission project, aiming at fully exploiting the unique capabilities of GRBs for cosmology and multi-messenger astrophysics.

In frames of the program optical observations
with the 6m BTA telescope of SAO RAS~\citep{Sokolov2018a,Vlasyuk2018} are
conducted. We look for fast variability of optical
flux of GRB afterglows both in imaging and spectroscopic
modes.
Using the BTA MANIA fast photometry facility one can
detect  very short optical variability ($\tau = R/c$).
Thus these observations are especially important for testing
for testing strong regime of gravity theory, because it determines the size of a RCO. Observed polarization and possible RCO surface effects (magnetic field, hot spots) can deliver crucial information on the RCO nature. 

\begin{figure}
    \centering
    \includegraphics[width=0.49\textwidth]{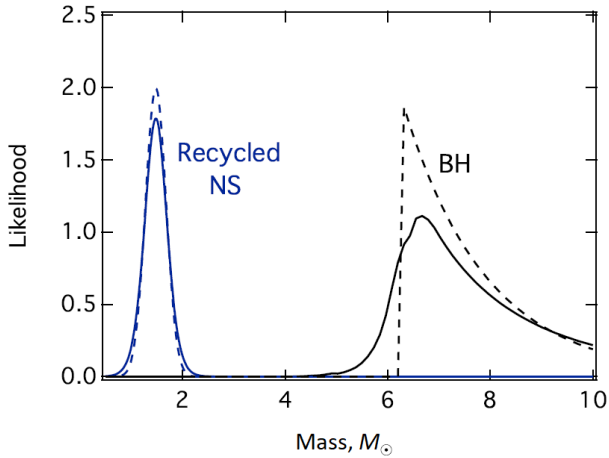}
    \caption{The inferred mass distributions for the different populations of neutron stars (left) and black hole candidates (right) discussed in the~\citet{Ozel2012}. The dashed lines correspond to the most likely values of the parameters. For the recycled neutron star populations the peak is $M_0 = 1.48M_{\odot}$ and $\sigma = 0.20M_{\odot}$. For the case of black hole candidates the peak is $M_\text{c} = 6.32M_{\odot}$ and a scale of $M_\text{scale} = 1.61M_{\odot}$. The solid lines represent the weighted mass distributions for each population.}
    \label{fig:NS_BH_population}
\end{figure}

In particular, general relativity predicts black holes of a mass of $M > 3\,M_{\odot}$ for RCO, while  the quantum field approach to gravitation predicts a critical mass of 6.7 $M_{\odot}$ for a quark RCO
\citep{Sokolov2015,Sokolov2016,Sokolov2019}.
Fig.\ref{fig:NS_BH_population} presents an ``unexplained''
observed mass gap $2-5\,\,M_{\odot}$ between neutron stars and black hole candidates
\citep{Ozel2012}. The nature of the RCO with a mass of 
$M>5\,M_{\odot}$ is the laboratory for crucial testing of the gravity theory.

Up to now,  GW170817/GRB170817A is the only  observation of gravitational waves originating from the merging of two compact objects in the mass range of neutron stars, accompanied by electromagnetic counterparts, and offers an opportunity 
to probe the internal structure of neutron stars directly
\citep{Abbott2017}.
These observations support the hypothesis that GW170817 was produced by the merger of two neutron stars in
NGC4993 followed by a short GRB 170817A.
Future THESEUS--BTA observations of GRBs will essentially increase statistics of such crucial events and hence make crucial contribution to testing gravitation theory as the basis of cosmological models.

\section{Testing Cosmological Principle by GRB observations}

The sources of GRBs are massive supernovae explosions and merging of binary RCOs~\citep{Sokolov2018a}.
They mark their host galaxies up to high redshifts, hence their observations can probe the spatial large-scale distribution of visible matter.

\paragraph{Large-scale distribution of galaxies.}
Modern progress in spectral and photometric redshift surveys for wide-angle (e.g., 2dF, SDSS, BOSS) and deep fields (e.g., COSMOS)
leads to the discovery of very large structures at all observed redshifts. Direct observations of the
spatial distribution of visible matter (galaxies) do reveal  inhomogeneity on scales much larger than the standard Peebles's correlation length $r_0 \sim 5$ Mpc.

Nowadays the observationally established scales of inhomogeneity reach  several hundreds Mpc. 
The Laniakea supercluster of galaxies \citep{Tully2014}
and the Dipole Repeller with the Shapley Attractor \citep{Hoffman2017} in the Local Universe reach a size of
$\sim$200 Mpc.
The Sloan Great Wall has a size of $\sim$100 Mpc at a distance of $\sim$200 Mpc~\citep{Einasto2016,Gott2005}. 
The SDSS/CMASS survey discovered the BOSS Great Wall with a size of $\sim$300 Mpc at a distance of $d \sim 2\,000$ Mpc~\citep{Lietzen2016}.
In the ultra deep galactic field (UDHF) the photometric redshift survey COSMOS revealed evidence for Super Large Clusters with sizes of $\sim$1\,000 Mpc at $z \sim 1$ \citep{Nabokov2010b,Shirokov2016}.

\paragraph{Large-scale distribution of GRBs.}
Studies of spatial distribution of GRBs with known redshifts also revealed very large inhomogeneous structures, though with large uncertainty.
A giant ring of GRBs with a diameter of 1\,720 Mpc at redshifts of $0.78 < z < 0.86$ has been found in \citet{Balazs2015}. 
The probability of observing such a ring-shape structure by chance
is $2 \times 10^{-6}$.

The spatial distribution of 244 GRBs has been analyzed as part of the Swift mission using the Peebles $\xi$-function method by
\citet{Li2015}. 
They obtained the correlation length  $r_0 \approx 388\,h^{-1}$ Mpc, 
$\gamma = 1.57 \pm 0.65 $ (at the 1$\sigma$
level), and the uniformity scale is  $r \approx 7700\, h^{-1}$.

These facts require reconsideration of  
the basic $\Lambda$CDM principles of homogeneity and isotropy distribution of matter and it evolution with cosmic time.

\subsection{Fractal properties of spatial distribution of GRBs}

In general physics fractal structures naturally originate in phase transitions, dynamical chaos, strange attractors and other physical phenomena. Fractals are characterized by the power-law correlations in a wide range of scales.

The fractal model of spatial distribution of galaxies with the fractal dimension close to the critical value $D=2$ finely describes the data of many redshift surveys~\citet{Gabrielli2005,Baryshev2012}.

As it was demonstarted by 
\citet{Gabrielli2005,Baryshev2012} the Peebles`s reduced correlation function~\citep{Peebles1993} $\xi(r)$ is strongly distorted by the borders of real samples. To get robust statistical characteristics of the spatial distribution of galaxies one should use the complete correlation function, called also the conditional density function $\Gamma(r)$. In particular, for a fractal spatial distribution the slope of power-law $\Gamma(r) \propto r^{-\gamma}$ gives the robust estimation of the fractal dimension $D=3-\gamma$ (for the homogeneous  distribution $\gamma=0$ and $D = 3$).

The conditional density and pairwise distances as methods of fractal analysis were proposed
by \citet{Grassberger1983a, Grassberger1983b}.
Conditional density analysis of main galaxy samples was 
developed in 
\citet{Gabrielli2005,Baryshev2012,SylosLabini2014}.
The pairwise method was developed in~\citet{Raikov2011, Shirokov2017}. 

In papers~\citet{Gerasim2015,Raikov2010} the fractal dimension of a GRB sample was estimated by the method of pairwise distances. They derived values of the fractal dimensions in the interval  $D = 2.2 \div 2.7$,  but only on scales up to 50 Mpc. 

In paper \citet{Shirokov2017}  the new modified methods of conditional density and pairwise distances were presented, which allow one to estimate the fractal dimension at the full interval of  scales for a given sample.
The normalized distributions of the conditional density and pairwise distances for real GRB sample and for fractal model catalogs give values of the fractal dimension $D \approx 2.0$ and 
$D \approx 2.5$ respectively. 
For the case of a full celestial sphere, the conditional density method gives the fractal dimension of distribution of GRB sources  equal to $D = 2.6 \pm 0.12$ at $r = 1.5 \div 2.5$ Gpc  and $D = 2.6\pm0.06$ for $r = 1.5\div5.5$ Gpc.  The pairwise distances method gives a stable power law dependence with $D = 2.6 \pm 0.06$ and  does not change essentially for the interval of linear scales $l = 1.5 \div 5.5$. Thus, on scales of $\approx 1.5 \div 5$ Gpc, both methods of GRB spatial structure analysis give a similar exponent of the power law correlation. However the number of GRBs with measured redshifts in analyzed samples is still too small ($N<300$), and the above estimations  are preliminary results.

\subsection{Isotropy of distribution of GRBs}
 
Isotropy of the distribution of GRBs in the celestial sphere by the {\it Fermi}, {\it BATSE} and {\it Swift} data was analyzed in paper~\citet{Ripa2018}. Authors considered the observed properties of GRBs and made the conclusion: ``...the results are consistent with isotropy confirming''.

However, 
anisotropy of sky distribution of GRBs was detected in a number of papers
\citet{Balazs2015,Raikov2010,Gerasim2015, Shirokov2017}.
Thus, for example, a spatially
isolated group of five GRB was detected with the coordinates
$23^h 50^m < \alpha < 0^h 50^m$ and $5^0 < \delta < 25^0$ at redshift of $0.81 < z < 0.97$ and also they
found  GRB groups in several directions on the sky.

It should be emphasized that homogeneity and isotropy of spatial distribution are different properties of the large-scale structure 
\citep{Gabrielli2005,Baryshev2012,SylosLabini2014}. 
For example the fractal distribution of matter can have statistical isotropy and simultaneously be strongly inhomogeneous and the Copernican Principle is fulfilled  \citet{SylosLabini2010}.  
Our mentioned results on  fractal dimension $D$ being close to it critical value $D_{crit}=2$ on a very large interval of scales demonstrate that such a situation may be realized in the spatial distribution of GRBs.

Future THESEUS--BTA observations will essentially increase the number of GRBs with known redshifts and hence allow one to get strong restrictions on Cosmological Principle of homogeneity and isotropy of visible and dark spatial (and line-of-sight) distribution of matter.

\subsection{Cosmic tomography via line-of-sight observations of GRBs}

\paragraph{Deep pencil-beam galaxy survey.}
An important goal of cosmology is to set an observational limit on the sizes of the largest structures in visible distribution of galaxies. Recent deep spectral and multi-band photometric surveys of galaxies 
give a new possibility to estimate a homogeneity scale on which the  distribution of luminous matter becomes uniform.

Statistical analysis of the number density fluctuation for the various pencil-beam deep galaxy surveys (COSMOS, HUDF, ALHAMBRA) was considered in \citet{Nabokov2010a,Nabokov2010b,Shirokov2016}.

Observational ``cosmic tomography'' test on the reality of the super-large structures (having large angular size on the sky) was suggested in \citet{Nabokov2010b}. 
It can be made possible by a lot of narrow angle (a few arc-minutes) very deep multi-band photometric beam-surveys in the grid nodes covering the sky (the cells are a few degrees and more).
Then, increasing the number of nodes of the grid one can probe the extension of the super-large structure in tangential direction. An advantage of this method is that the very deep faint galaxy surveys allow one to achieve the very wide-angular extensions needed for observations of super-large structures. 

\paragraph{THESEUS--BTA Cosmic Tomography.}
Important application of this method is to consider the directions to GRBs as nodes. In this case, deep-field observations, which are needed for observations of the GRB host galaxies,  play the role of nodes of the grid~\citep{Baryshev2010, SokolovJr2016}.

As a result of such observational program one can construct the 3D map of  ``super-structures'' by performing correlation between neighboring radial redshift distributions,  i.e. to perform the ``cosmic tomography'' of the observable Universe. 

\begin{figure}
    \centering
    \includegraphics[width=0.48\textwidth]{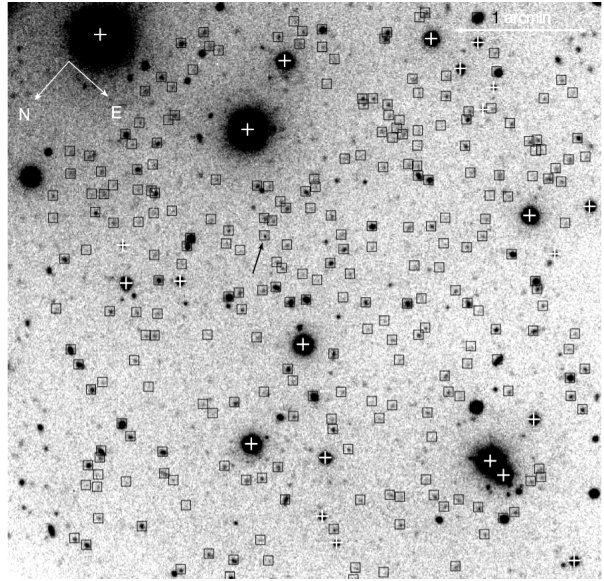}
    \caption{The objects detected in four filters (the galaxies are enclosed by the squares, the star-shaped objects are marked by crosses). The black arrow points to the host galaxy of GRB021004.}
    \label{fig:tomoBTA}
\end{figure}

\begin{figure}
    \centering
    \includegraphics[width=0.48\textwidth]{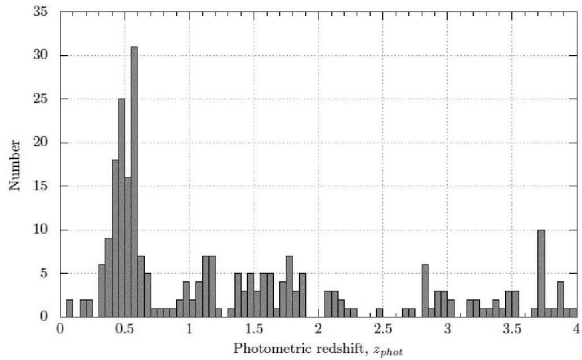}
    \caption{The photometric redshift distribution for 246 objects with the peak at $z \approx 0.56$ based on BTA $BVRI$ data.}
    \label{fig:dN(dz)}
\end{figure}

An example of such a node of a possible future grid is the BTA deep field of GRB021004.
In Fig.~\ref{fig:tomoBTA} the direct BTA image centered on the GRB021004 host galaxy 
is shown
\citep{Baryshev2010, SokolovJr2016, SokolovJr2018}. The photometric redshift is measured for each faint galaxy in the field and the number of galaxies $dN(z)$ in the redshift bin $dz$ is presented in Fig.~\ref{fig:dN(dz)}.
The smoothed peaks correspond to the galaxy clusters along the GRB line-of-sight.

The observed distribution of galaxies along the line-of-sight gives information about inhomogeneous distribution of visible matter in the fixed direction in the sky. Statistical analysis of a grid of such fields will allow one to perform a tomography of the large-scale distribution of galaxies on largest optically available scales
~\citep{Nabokov2010b, SokolovJr2016, Shirokov2016,SokolovJr2018}.

This test may be executed step by step by deep BTA observations of neighboring GRB host galaxies, close to an initial GRB direction.

In a sense, in the beginning of the 20$th$ century, using the largest at that time telescopes, Edwin Hubble opened the door into the ``realm of galaxies'', and now, in the beginning of the 21st century, by operating with 
multimessenger gamma-optical facilities, we have an opportunity of the observational studying the ``realm of metagalaxies''.

\section{Testing the space expansion paradigm}

The most important cosmological observational fact, which was discoverer by \citet{Hubble1929}, is the linear (for small distances) relation between observed redshift $z$ of the spectral lines and the distance $r$ to a galaxy,
i.e. the observed Hubble Law (redshift--distance relation):
\begin{equation}
\label{z-r-law}
z(r) = \frac{H_0\,r}{c} = \frac{r}{R_H}
= \frac{V_{app}(r)}{c}  \,\,,
\end{equation}
where $H_0$ is the Hubble constant at present time,
$R_H = c/H_0 \approx 4000\,h^{-1}_{75}$ Mpc is the Hubble radius, $V_{app}= c\, d\lambda/\lambda$ is the apparent spectroscopic radial velocity of a galaxy, $c$ is the velocity of light,
\begin{equation}
\label{z-law}
z = \frac{\lambda_{obs} - \lambda_{emit}}{\lambda_{emit}}    =\frac{\lambda_{obs} }{\lambda_{emit}} - 1 
= \frac{V_{app}(r)}{c}\,\,,
\end{equation}
$\lambda_{obs}$ is the observed photon wavelength  at a telescope and $\lambda_{emit}$ is the wavelength of emitted photon at the distance $r$  in the observed galaxy.

To determine the distance $r$ to a galaxy Hubble used the concept of the ``standard candle'', i.e. an object with an apriori known luminosity.
Note that Hubble called the redshift ``an apparent velocity'' (in units of $c$) because he measured the distance $r$ to a galaxy through the flux measure, and he did not measure the physical velocity of a galaxy as the change of distance with time.

The expanding space paradigm of the Standard Cosmological Model is the theoretical interpretation of the redshift as the Lema\^{i}tre effect in the expanding Friedmann universe where the space expansion velocity is
$V_{exp}(r) \equiv V_{app}(r)$.

Surprisingly, after almost hundred years after
Lema\^{i}tre's interpretation of cosmological redshifts as effect of the space expansion, the acute discussion
again raised in professional cosmological literature
about physical sense of the  cosmological redshift and 
relation between mathematical geometrical concepts and measured astronomical quantities: 
\citet{Harrison1993, Harrison1995, Harrison2000,Peacock1999, Peacock2008,Davis2004, Davis2010,Abramowicz2007, Abramowicz2009,Kaiser2014,Baryshev2015}.

\subsection{Theoretical Hubble Diagram for different cosmological models}

Cosmological Hubble Diagram (HD) incorporates
the directly observed fluxes, luminocity distances and redshifts for a particular class of standard candles. This is why the HD can be used for observational testing the basic theoretical relations of cosmological models.

\paragraph{The Standard Friedmann's Models.}
The expanding space paradigm states that the proper (internal) metric distance $r$ to a galaxy, having fixed comoving coordinate $\chi$ from the observer, is given by the relation 
\begin{equation}
\label{space-exp}
r(t) = S(t)\cdot \chi    
\end{equation}
and increases with time $t$ as the scale factor $S(t)$.

It is important to point out that the hypothesis of homogeneity and isotropy of space (Cosmological Principle)  implies that for a given galaxy the recession velocity is proportional to distance via \textit{exact linear} velocity--distance 
($V_{exp}$ vs $\,r$) relation 
for all FLRW metrics 
\begin{equation}
\label{expvel}
    \begin{gathered}
\frac{V_{exp}(r)}{c} = \frac{dr}{cdt} =  
\frac{dS}{cdt} \chi = 
    \frac{H(t)\, r}{c} =  \frac{r}{R_{H}},
    \end{gathered}
\end{equation}
where $H=\dot{S}/S$ is the Hubble parameter  and $\,R_{H} = c/H(t)$ is the Hubble distance at the time $t$. Note that from Eq.(\ref{expvel})
one gets expansion velocity more than velocity of light $V_{exp}(r) > c$ for $\,r> R_{H} $
\citep{Harrison1993, Harrison2000,Baryshev2012} .

It should be emphasized that the cosmological expansion velocity  $V_{exp}(r)$ for an observed galaxy is conceptually different from the galaxy peculiar velocity
$V_{pec}$, which can not be larger than the velocity of light. 
The cosmological redshift in expanding space is not the Doppler effect, but the Lema\^{i}tre effect defined as the ratio of scale factors: 
\begin{equation}
\label{lem}
 (1+z) = \frac{\lambda_{0}}{\lambda_{1}} =
 \frac{S_{0}}{S_{1}}\,,    
\end{equation}

Instead of exact linear relation Eq.(\ref{expvel})
for $V_{exp}(r)$, the redshift--distance relation $z(r)$ and expansion velocity--redshift relation $V_{exp}(z)$  are non-linear:
\begin{equation}
\frac{V_{exp}(z)}{c}= \frac{r(z)}{R_{Ho}} = 
 \int_0^z\frac{dz'}{h(z')}\,,   
\end{equation}
where $h(z) = H(z)/H_0$.

\paragraph{The SCM Magnitude -- redshift relations.}
When observed through the ``i''  filter an object with the absolute magnitude $M_i$ has the apparent magnitude $m_i$,
\begin{equation}
    m_i(z)=5\log(l(z)(1+z))+C_i(z),
    \label{m-LCDM}
\end{equation}
where $C_i(z) = 25 + M_i + K_i(z) + A_i + E_i(z)$, 
and $l(z)$ is the external metric distance 
\citep{Baryshev2012}: $l(r)=S(t)I_k(r/S)$.

If the K-correction, extinction, and evolution corrections are known for a standard candle class, Eq.~(\ref{m-LCDM}) can be used to derive the redshift--luminosity distance relation $l_{lum}(z) = l(z)(1 + z)$.
The ``pure vacuum'' flat model ($\Omega = \Omega_\Lambda = 1$) has the linear relation $l(z) = r(z) = R_{H_0}z$, hence
$m_i(z) = 5 \log(R_{Ho}\,z(1 +z)) + C_i(z)$.

\paragraph{The Classical Steady State Model.}
The luminosity and metric distances are related similarly
as in the Friedmann flat models: $r(z) = l_{lum}(z)/(1+z)$. Because in CSSM $r(z) = R_H z$, the magnitude--redshift relation is
\begin{equation}
    m_i(z) = 5 \log(R_H z(1 + z)) +C_i(z).
    \label{m-SS}
\end{equation}

\paragraph{The Fractal Cosmological Model.}
In the framework of the fractal cosmological model by
\citet{Baryshev2008,Baryshev2012,Baryshev2017}
the Universe is isotropic and inhomogeneous with a fractal dimension $D \approx 2.0$. Such a value of the fractal dimension guaranties the linear redshift--distance law, if the cosmological redshift is interpreted as the global gravitational redshift within the fractal structure.

In the fractal model, the luminosity and metric distances are related as $r_{lum}(z) = r(z)(1 + z)$. This result includes the lost energy of individual photons
and their diminished arrival rate due to gravitational time dilation. The magnitude--redshift relation then becomes
\begin{equation} \label{FF}
    m_i(z) = 5 \log(R_H Y(z)(1 +z)) +C_i(z),
\end{equation}
\noindent where function Y(z) is defined in~\citet{Baryshev2008,Baryshev2012}.

\paragraph{The Zwicky Tired-Light Model.}
The Zwicky TL model can be used as a toy example,
where  cosmological time dilation effect is excluded.
In the simplest tired light model with the Euclidean static space (for instance, La~Violette (1986)~\citet{LaViolette1986}) the magnitude of a standard candle depends on the redshift as follows:
\begin{equation} \label{TL}
    \begin{gathered}
       m_i(z) = 5 \log(R_H ln(1 +z)) + \\ 
       + 2.5 \log(1 +z) + C_i(z).
    \end{gathered}
\end{equation}


\subsection{Observed Hubble Diagram for high-redshift long GRBs}

\paragraph{Construction of Hubble Diagram of the long GRBs.}
The Hubble Diagram (HD) is the directly observed relation between flux and redshift
for a sample of standard candles. The Hubble Diagram test is expressed as the distance modulus versus redshift for ``standard''  GRBs, calibrated by means of the ``Amati relation'':
\begin{equation}
    \mu = 5 \log {\frac{d_L}{Mpc}} + 25 ,   
\end{equation}
\noindent where $\mu$ is the distance modulus and $d_L$ is luminosity distance. The latter is given by
\begin{equation}
    d_L = \left( \frac{(1+z)E_{iso}(E_{p,i})}{4\pi S_{bolo}} \right)^{\frac{1}{2}}, 
\end{equation}

\noindent where $S_{bolo}$ is GRB fluence and $E_{iso}(E_{p,i})$ is the isotropic energy calculated by the Amati relation $E_{p,i}=K\times E_{iso}^m$~\citet{Amati2009}. The cosmological rest-frame spectral peak energy $E_{p,i}=E_{p,obs}\times(1+z)$.

The HD method is based on application of theoretical relations given by Eqs.(\ref{m-LCDM}, 
\ref{m-SS}, \ref{FF}, \ref{TL}).
Detailed analysis of the GRB Hubble Diagram as a high-redshift cosmological test is presented in 
\citet{Shirokov2020}.

\begin{figure}
    \includegraphics[width=0.49\textwidth]{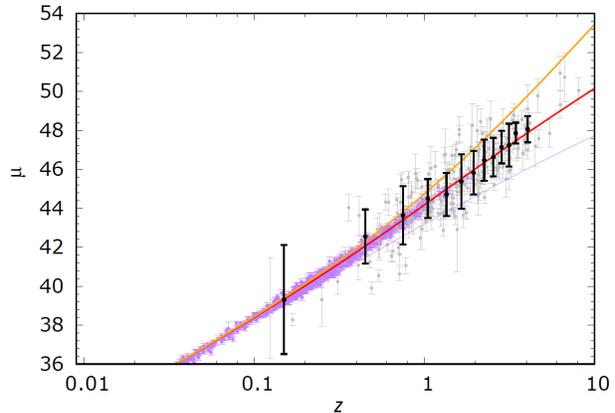}
    \caption{The Hubble Diagram for the SNe Ia Pantheon catalog from~\citet{Scolnic2018} (purple points), the GRBs catalog from~\citet{Amati2019} (gray point), the median values of GRB bins with $\Delta z=0.3$ (black points), and the predictions of $\Lambda$CDM model (red curve) and two  examples of other models (pure vacuum and no time dilation).}
    \label{fig:HD}
\end{figure}

As an example of the modern HD testing of the basic cosmological relations $z(r),\, F(z)$,
we take the distance modulus of a sample of 193 long GRBs from~\citet{Amati2019} and calculate the median distance modulus values in redshift bins. In Fig.\ref{fig:HD} purple points are the SNe Ia Pantheon catalog from~\citet{Scolnic2018}, gray points are GRBs with known redshifts and spectra from paper~\citet{Amati2019}, black points are the median values in redshift bins with $\Delta z=0.3$ (black points), the red curve is the prediction of the $\Lambda$CDM model with $\Omega_\Lambda=0.7$, the orange curve is ``pure vacuum'' Friedmann's model and purple curve is the no time dilation Zwicky tired-light model.

Figure~\ref{fig:HD} shows that  predictions of different models have several stellar magnitudes
in the THESEUS redshift interval (up to $z\sim 10$).
So the future THESEUS observations will give essential extension and accuracy of the observed Hubble law and hence put strong new cosmological model restrictions.

\paragraph{Gravitational lensing and Malmquist biases in the Hubble Diagram.}
There are a lot of observational selection effects (e.g. limits on detector sensitivity, influence of  interfering matter, gravitational lensing, beaming effect, evolution), which potentially distort the measured bolometric flux and fluence,  and hence the derived distance to a GRB.
Thus to the construction of the intrinsic Hubble Diagram one should take into account the selection effects. However, a firm answer to this fundamental question is far from being settled until more GRB data with known redshifts are available.

The gravitational lensing of long GRBs by gravitating matter, located along the ``source--observer''  line of sight, produces  apparent increase of flux and fluence $S_\text{bolo}$ due to gravitational lens magnification, which does not change frequency (and $E_\text{p}$) of the lensed radiation. This can be misinterpreted as an evolution of GRB luminosity.

According to~\citet{Ji2018}, GRBs can be magnified by the gravitational lensing produced by different gravitating  structures of the Universe (such as dark and luminous stellar mass objects, globular and dark stellar mass clusters, galaxies and dark galaxy mass objects). 
Hence the gravitational lensing can have a great impact on high-redshift long GRBs. For example, according to~\citet{Kurt2000, Ougolnikov2001}, 
in the BATSE catalog there are several GRBs which are lensed by the intergalactic globular clusters.

If we take into account that there is a threshold for detection in the burst apparent brightness, then, with gravitational lensing, bursts just below this threshold might be magnified in brightness and detected, whereas bursts
just beyond this threshold might be reduced in brightness and excluded (the Malmquist bias). 

As it is demonstrated in \citet{Shirokov2020}, the combined gravitational lensing and Malmquist biases crucially influence on the observed Hubble Diagram.
If one takes into account possible luminosity correction on high-redshift GRB HD, then the observed HD tends to be consistent with the $\Lambda$CDM having the vacuum density parameter $\Omega_L \rightarrow 0.9$ and the dark matter density parameter  $\Omega_m \rightarrow 0.1$.
This result is very important in view of recent discussion about the role of dark energy and dark matter
\citet{Makarov2011,Colin2019,Demianski2019,DiValentino2020}.

So the crucially important fundamental question on the role of the gravitational lensing bias in high-redshift long GRB data needs more observational and theoretical studies. 
In  particular, the 6m BTA observations  \citet{SokolovJr2016,SokolovJr2018}
of galaxies along the long GRB line of sight Fig.\ref{fig:dN(dz)}  will be important for estimation of the lensing magnification probability of the long GRB fluxes and it influence on the high-redshift HD. 
Hence the THESEUS--BTA joint program will give crucial information on dark matter and dark energy. 

\subsection{Time Dilation Cosmological Test}

One of the crucial cosmological tests on the nature of the cosmological redshift is the measurements of duration of known physical processes at high-redshift objects (cosmological time dilation). 

\paragraph{Wilson’s Supernova Time Dilation Test.}
\citet{Wilson1939} suggested supernovae as a test of the nature of the cosmological redshift: in an expanding universe the light-curve of a supernova occurring in a distant galaxy should appear to be expanded along the time axes in the ratio $(1 + z) : 1$ with respect to the standard local light-curve. This time delay test was also discussed by \citet{Rust1974} and \citet{Teerikorpi1981}. 

Recent observations of the Ia supernovae have finally given an opportunity to perform the test \citep{Leibundgut2001,Goldhaber2001}.
The observed width $\tau_{obs}$ of the supernova light-curve can be written as
\begin{equation}
    \tau_{obs}(z) = \tau_{em}(1 +z)^p,
\end{equation}
where $p = 1$ for the local Doppler and gravitational effects, and also for Lema\^{i}tre space expansion effect and de Sitter-Bondi global gravitational effects, while $p = 0$ for all models without cosmological time dilation. 

Light curves for 35 Type Ia supernovae with redshifts up to $z \approx 1$ were analyzed by \citet{Goldhaber2001}. They derived the dilation parameter $p = 1.0\pm0.1$. The spectral ages in the supernova rest frame were measured in another study by of~\citet{Blondin2008}. Comparison with the observed time led again to the $(1 + z)^{1}$ factor expected for expanding space and also gravitational nature of cosmological redshift.

Note that the time dilation test provides good evidence against the tired-light hypothesis, but it can not distinguish between expanding space models and involving cosmological global gravitational redshift models. 

\paragraph{GRB Pulse Stretching Test.}
As an  observational test of the time dilation effect one can consider the relation $T_{90}\propto(1+z)^\alpha$ for GRB pulse profiles. \citet{Kocevski2013} and \citet{Zhang2013} considered the dependence of GRB pulses duration on redshift and got the conclusion that the slope is about $\alpha \approx 1$. A similar result $T\propto(1+z)^{1.4\pm0.3}$ has been obtained for the radio loud GRBs sample in the paper~\citet{LloydRonning2019}. 

The time dilation test can be made separately for Long and Short GRBs.
The long GRBs are explosions of supernovas, while the short GRBs are mergers of binary systems. These events have the same physical nature as a result of relativistic gravitational collapse, but  different light curves due to matter envelop \citep{Sokolov2018a,Dado2018}. GRBs are usually divided into long-soft ($T_{90} < 2$s) and short-hard ($T_{90} < 2$s), which are less than 10\%, e.g., in the {\it Swift} sample~\citet{Shirokov2019}. 

In fact the observational selection effects can strongly influence the observed duration of GRBs at different redshifts. In particular,
it strongly depends on instrumental time resolution and spectral sensitivity, and also on spectral features of sources~\citep{CastroTirado2018, Kocevski2013}. Hence the time dilation test  is hard to perform by GRB pulses alone.

\paragraph{New GRB+SN cosmological test of time dilation.}
Time dilation of all physical processes observed at high-redshift is  predicted by both cosmological models based on space expansion and on global gravitational redshift mechanisms.
At small redshifts it is difficult to measure this effect
because of  different other competitive physical processes.

A new opportunity exists for Wilson cosmological test of the time dilation will appear, when THESEUS GRB gamma-ray pulses observations will be used together with the same GRB afterglows light curves observed with the 6m BTA SAO  telescope
\citep{Amati2018,Vlasyuk2018,Sokolov2018a}.

Because long GRBs are related to core-collapse SN explosions, there is a  possibility for  cosmological test of the time dilation effect. One can consider simultaneously the shape of gamma-ray pulses and the shape of an afterglow in different bands. For example, the SN light curve  can be visible in the GRB afterglow light curve.
Hence time duration of different processes can be used  for the same GRB, and their statistics in different channels and different redshifts will present the robust estimations of the time dilation effect. 
A cooperative THESEUS--BTA observations will be important for this cosmological test of the fundamental physics.

\section{Conclusions}

In the spirit of the Sandage's practical cosmology approach  \citep{Sandage1997,Sandage1995a},
we have considered the current state of cosmology, which is characterized by general tendency to testing the fundamental principles lying in the basis of cosmological models. The especially important role belongs to the recent discovery of a discrepancy  between Planck-2018 results on CMBR fluctuations analysis and the locally measured cosmological parameters of the SCM. 

Such obstacles as the nature and value of dark energy and dark matter, the value of gravitational lensing by the large-scale structure and the value of the Hubble constant $H_0$ for the Local Universe, are now discussed as a new crisis for cosmology (see references in Introduction).

The GRB observations in multimessenger astronomy epoch open new possibilities for testing the fundamental physics lying in the basis of the standard cosmological model: classical general relativity, cosmological principle of matter homogeneity, and the Lema\^{i}tre space expansion nature of cosmological redshift. 
As Harrison also says: ``The history of cosmology shows that in every age devout people believe that they have at last discovered the true nature of the Universe''~\citet{Harrison2000}.

Modern achievements of the theoretical physics,  especially  different modifications of general relativity and  quantum aspects of gravitation theory, together with a number of conceptual problems of the SCM, also require a reanalysis and the wider observational testing  of the initial principles of the SCM.

We have considered  possible basic cosmological applications of GRBs multimessenger observations in the wide interval of cosmological redshifts up to $z \sim 10$. THESEUS--BTA cosmological tests can probe strong-field regime of gravitation theory,  spatial distribution of galaxies, Hubble Law  and time dilation of physical processes at such redshifts.
Perspectives for performing these cosmological tests in multimessenger astronomical observations of GRBs were considered and several new tests were proposed. The very important part of cosmological tests is related to careful taking into account different selection observational effects that distort the true cosmological relations.

Future THESEUS space observations of GRBs~\citep{Amati2018, Strata2018} and corresponding multimessenger ground-based  studies, including large  optical telescopes, such as BTA and GTC, and even 1-m class telescopes \citep{Vlasyuk2018,CastroTirado2018, SokolovJr2016, Sokolov2018a}, will bring crucial information for testing theoretical cosmological models. 

Our analysis of possible application of  observational cosmological tests for forthcoming THESEUS GRB mission has demonstrated its potentially fundamental contribution to cosmology, because GRBs are among the most distant astrophysical objects with measured spectral redshifts (see Fig.~\ref{fig:HD}).

The promising cosmological tests of the SCM basis by forthcoming THESEUS and BTA observations of GRBs are:
\begin{itemize}
    \item BTA identification and monitoring of fast optical counterparts for THESEUS GRBs, together with detections of neutrino and gravitational wave signals, allows one to test the strong regime of gravity theory as the basis of the standard cosmological model. The problem of transition of relativistic compact objects to neutron stars, quark stars or black holes can be solved and extended to cosmological solutions of the gravity field equations.
     \item large number of the THESEUS GRBs and BTA optical observations of host galaxies make it possible to test the Cosmological Principle of homogeneity and isotropy on largest spatial scales  up to $r\sim 10$ Gpc. The ``cosmic tomography'' of the large-scale structure can be studied by using  BTA deep-field observations of line-of-sight distribution of galaxies in directions of GRBs.
    \item THESEUS high-redshift Hubble Diagram for long  GRBs  in collaboration with BTA line-of-sight observations, allows one to test the flux--distance--redshift cosmological relations  up to $z \sim 10$. The joint THESEUS--BTA observations of the gamma-Xray-optical-IR light curves profiles  will give an opportunity to perform a new form of the Wilson's time dilation test  for high-redshift physical processes.
\end{itemize}

The joint THESEUS--BTA GRB project, together with  other multimessenger observations, will give decisive a new information on the fundamental cosmological physics. 

\begin{acknowledgements} 
We are grateful to O. V. Verhodanov, D. I. Nagirner, and A. J. Castro-Tirado for useful discussions and comments. We thank T. N. Sokolova for editing corrections. 
The work was performed as part of the government contract of the SAO RAS approved by the Ministry of Science and Higher Education of the Russian Federation.  
\end{acknowledgements}

\bibliographystyle{my_arXiv}
\bibliography{references_A.bib}

\end{document}

%% file: cmd_arXiv.tex
\def\squareforqed{\hbox{\rlap{$\sqcap$}$\sqcup$}}

\def\sq{\ifmmode\squareforqed\else{\unskip\nobreak\hfil
\penalty50\hskip1em\null\nobreak\hfil\squareforqed
\parfillskip=0pt\finalhyphendemerits=0\endgraf}\fi}

\def\utw{\smash{\rlap{\lower5pt\hbox{$\sim$}}}}

\def\udtw{\smash{\rlap{\lower6pt\hbox{$\approx$}}}}

\def\diameter{{\ifmmode\mathchoice
{\ooalign{\hfil\hbox{$\displaystyle/$}\hfil\crcr
{\hbox{$\displaystyle\mathchar"20D$}}}}
{\ooalign{\hfil\hbox{$\textstyle/$}\hfil\crcr
{\hbox{$\textstyle\mathchar"20D$}}}}
{\ooalign{\hfil\hbox{$\scriptstyle/$}\hfil\crcr
{\hbox{$\scriptstyle\mathchar"20D$}}}}
{\ooalign{\hfil\hbox{$\scriptscriptstyle/$}\hfil\crcr
{\hbox{$\scriptscriptstyle\mathchar"20D$}}}}
\else{\ooalign{\hfil/\hfil\crcr\mathhexbox20D}}%
\fi}}

\newcommand{\ab}{Astrophysical Bulletin}

\newcommand{\aaa}{Astron. and Astrophys.}
\newcommand{\aap}{Astron. and Astrophys.}

\newcommand{\aj}{Astron.~J.}

\newcommand{\araa}{Annual Rev. Astron. Astrophys.}

\newcommand{\mnras}{Monthly Notices Royal Astron. Soc.}
\newcommand{\pasa}{Publ. Astron. Soc. Australia}

\newcommand{\advspres}{Adv. Space Res.}

\newcommand{\astrep}{Astron. Rep.}
\newcommand{\cosmres}{Cosmic Res.}

\newcommand{\intjmodphysa}{Int. J. Modern Phys. A}
\newcommand{\livrevrel}{Living Rev. Relativ.}
 
\newcommand{\revmodphys}{Rev. Mod. Phys.}
\newcommand{\pnas}{Proc. Natl. Acad. Sci.}

\newcommand {\actc} {Acta Cosmol.}
\newcommand {\acta} {Acta Astr.}

\newcommand {\apjl} {Astrophys. J. Lett.}
\newcommand {\afz}  {Astrophysics}

\newcommand {\jcap} {J. Cosmol. Astropart. Phys.}
\newcommand {\lrr}  {Liv. Rev. Rel.}
\newcommand {\na}   {Nat. Astron.}
\newcommand {\nrp}  {Nature Rev. Phys.}
\newcommand {\nar}  {New Astron. Rev.}

\newcommand {\phu}  {Phys.-Usp.}

\newcommand{\physrep}  {Phys. Rep.}
\newcommand {\phrl} {Phys. Rev. Lett.}

\newcommand {\pdnp} {Physica D: Nonlinear Phenom.}
\newcommand {\sa}{Sci. Am.}